# Discovery Signal Design and Its Application to Peer-to-Peer Communications in OFDMA Cellular Networks

Michael. Wang, and C. Jiang

*Abstract*—**This paper proposes a unique discovery signal as an enabler of peer-to-peer (P2P) communication which overlays a cellular network and shares its resources. Applying P2P communication to cellular network has two key issues: 1. Conventional ad hoc P2P connections may be unstable since stringent resource and interference coordination is usually difficult to achieve for ad hoc P2P communications; 2. The large overhead required by P2P communication may offset its gain. We solve these two issues by using a special discovery signal to aid cellular network-supervised resource sharing and interference management between cellular and P2P connections. The discovery signal, which facilitates efficient neighbor discovery in a cellular system, consists of un-modulated tones transmitted on a sequence of OFDM symbols. This discovery signal not only possesses the properties of high power efficiency, high interference tolerance, and freedom from near-far effects, but also has minimal overhead. A practical discovery-signal-based P2P in an OFDMA cellular system is also proposed. Numerical results are presented which show the potential of improving local service and edge device performance in a cellular network.**

## I. INTRODUCTION

RAPID growth in the use of mobile computing devices, including smart phones and sensors, has triggered increasing demands for higher data rates and is challenging the conventional infrastructure of today's cellular communication systems. Today's cellular system is a typical centralized client-server based communication network. Any communication taking place has to go through base stations (i.e., the network access points). Devices do not interact directly in any manner with other devices in the network. This type of communication infrastructure is evidently not the most efficient, in terms of delay, system resource usage, power consumption and interference control, for local traffic generated by certain types of applications such as machine-to-machine (M2M) and gaming where a cluster of devices can be geographically close to each other and exchange messages amongst each other. Another critical drawback of the traditional cellular infrastructure that limits high data rate, high efficiency communication is the well-known cell edge effect. That is, device performance is highly dependent on the distance of the device to the base station (i.e., the geometry). A device that is far away from the base station (i.e., low geometry) suffers from poor performance or even outage due to insufficient link margin. The performance of cell edge devices is hence a bottleneck for cellular network performance.

The advent of self-organizing peer-to-peer (P2P) networks has improved interaction and collaboration among devices. P2P

communication, a form of collaborative communication, shows promising potential to resolve these critical issues and also to improve the scalability, fault-tolerance and heavy dependence on infrastructures of the centralized client-server based network model. P2P communication overlaying a cellular network has received much attention recently [1]-[6]. The integration of P2P communication into an LTE-Advanced network has also been investigated [7],[8]. These papers describe a peer-to-peer structure that is laid on top of the cellular infrastructure and shares the same radio resources with the cellular network. These high level studies show that, through cooperation and direct communications among devices, peer-to-peer capability increases user throughput performance. The results also show that P2P communication applied to cellular networks is a promising way to increase total network throughput, reduce power consumption, and improve coverage. It is therefore a perfect complement to today's cellular infrastructure.

While P2P communication has been shown at a high level to effectively address various problems exhibited in traditional cellular networks, many practical issues of adopting such a communication model have yet to be fully studied. In particular, excessive overhead could result from the signaling required by P2P for resource allocation and interference management. Such overhead, without careful management, can offset the gain from using P2P. The discovery signal (also termed a beacon) transmitted by the P2P devices for RF (radio frequency) proximity discovery is especially crucial to the operation of P2P since knowledge of RF proximity or neighbors is essential for P2P communications. Neighbor discovery, i.e., the detection of a device's immediate neighbors, is the first step in the process of P2P communications [9]-[11]. Due to the mobile nature of the devices in a cellular network, the topology of the network constantly changes. Even for certain devices of a static nature, connectivity is still subject to change even after the network has been established. The devices must look for neighbors on a regular basis to accommodate network topology changes. Interference tolerance and freedom from the near-far effect are the most challenging aspects in discovery signal design. Since multiple devices may simultaneously transmit their discovery signals, discovery signals should be designed in a way that multiple discovery signals do not interfere with each other; Hence discovery signals transmitted by closely located devices should not block the signals transmitted by devices farther away. Energy conservation is another factor that cannot be overlooked for most wireless devices. Since neighbor discovery is an on-going, periodic operation, the complexity and the energy efficiency of the discovery process has a large impact on the battery life of devices.

To meet the above challenging design goals, discovery signals usually require a large number of resources [12]-[16]. It is less of a problem in most ad hoc P2P networks since these networks typically operate in unlicensed bands. However, the high resource efficiency requirement in cellular networks, which operate in licensed bands, makes the traditional discovery signal design unsuitable for use in a cellular network.

The most popular discovery signals or beacons of IEEE 802.11/WiFi and IEEE 802.5/Zigbee systems consist of a preamble (a special sequence) followed by demodulation pilots and a message containing device information, and are transmitted on a block of contiguous frequency and time resources. The preamble field is composed of ten repetitions of a "short training sequence" for timing acquisition and coarse frequency acquisition and two repetitions of a "long training sequence" for channel estimation and fine frequency acquisition at the receiver. The beacon bandwidth ranges from 5 to 20 MHz corresponding to 52 subcarriers including four pilot subcarriers and 48 data subcarriers. The information contained in the beacon is QPSK-modulated and convolutionally coded just like a regular data frame except that a beacon frame is transmitted at a low data rate to ensure coverage [17]. The resulting beacon signal is therefore complex in the sense that it requires a preamble, pilots, and modulation/coding. The transmission of this "heavy-weight" signal not only requires many resources but also is sensitive to any type of co-channel transmissions (collisions).Therefore, this type of discovery signal is not suitable for cellular applications.

Another difficulty in applying conventional P2P to cellular networks is that resource and interference control is *distributed*. Tight interference and QoS (quality of service) control is thus hard to achieve without excessive signaling overhead. This presents a serious issue to a cellular network as providing stringent QoS is essential.

Therefore, the goal of this paper is to provide a mechanism to facilitate P2P communication overlaid on today's OFDMA cellular networks. Unlike existing designs, our design enforces the rule that P2P communications between devices be strictly supervised/scheduled by the network exactly as in the regular cellular connections except that the transmitter or the receiver is no longer the base station. This design is made possible via the use of a special discovery signal. The *key contribution* of this paper is the design (*at the physical layer level*) of a special low-overhead discovery signal that allows tight control of system resources by the cellular network in order to preserve the desired QoS requirements.

The rest of this paper is organized as follows. Section II gives a brief description of the basic neighbor discovery scheme providing baseline performance for neighbor discovery. Section III describes the proposed efficient neighbor discovery scheme. Section IV provides a concrete cellular P2P design example and simulation results to validate the proposed scheme. Section V concludes the paper.

## II. P2P IN CELLULAR NETWORKS

As mentioned earlier, the key to the proposed P2P structure is the use of the discovery signal in the enforcement of strict resource and interference control by the network. This network-supervised P2P communication in a cellular network is very similar to the traditional cellular communication (therefore minimum changes are required to the existing cellular networks) except that data are no longer necessarily relayed by the network. As depicted in Fig. 1, a source device first needs to find its peer device or target device to communicate with. If a direct P2P connection is a better choice to communicate with this device, the source device sends a request to the base station via uplink control channels. The base station signals an uplink grant to the P2P device pair via downlink control channels. The signaling/control messages between the device and the base station are via control channels just as in the regular cellular connection. The device pair uses the granted resource for traffic. However, the target receiver of the traffic sent on the uplink is no longer the base station but instead is the target device. The signaling (e.g., ACK/NACK) between the communicating device pair is via the cellular uplink control channels. However, the base station may monitor these messages to track the device status. In the P2P connection mode, since the base station no longer serves as a relay in the communication chain, downlink resources are thus saved for the regular cellular connection, which is the immediate benefit from using P2P (downlink is typically the most congested in today's cellular systems). As a result, the associated control channels for supporting the saved downlink traffic can now be freed. Another benefit, particularly for cell edge devices, of using P2P is that a device can use less power for higher data rate since P2P devices are close to each other. The data rate is no longer limited by the geometry of the device in the cell, i.e., low geometry devices can communicate with each other at high data rates. Using the example in Fig. 1, although devices A and B are far away from the serving base station, they still can enjoy high rate transmission which is not possible for traditional cellular operation as shown Fig. 2, where the data rate is ultimately limited by the uplink link budget.

In another scenario, illustrated in Fig. 3, devices A and B are far away from each other and both are far from the base station. In this case, the local traffic between A and B can be relayed by a middle device C and high data rate communication can still be established. In the case where traffic is not local, data have to be relayed by the cellular network in order to deliver the data from device D to device G located in cell Y that is far away from cell X. A direct link from device D to base station X does not sustain a high data rate. But device E can serve as a relay to base station X to sustain a higher uplink data rate. Similarly, the data that arrive at base station Y destined to device G can be relayed by device F to sustain a higher downlink data rate.

However, a question that follows is then: How should a device discover its target device or how should a device know its target device is in its neighborhood in an RF sense (i.e., RF proximity)? Also, how should the device find the best relay for a given target device? Neighbor peer discovery is indeed the key to the application of P2P to cellular networks and will be the focus of this paper. In particular, we look for a resource efficient, "light-weight" discovery signal design optimized for use in a cellular network.

## III. PEER DISCOVERY IN CELLULAR NETWORKS

Peer discovery for cellular networks can be, in principle, network-based or RF-based. In network-based discovery,

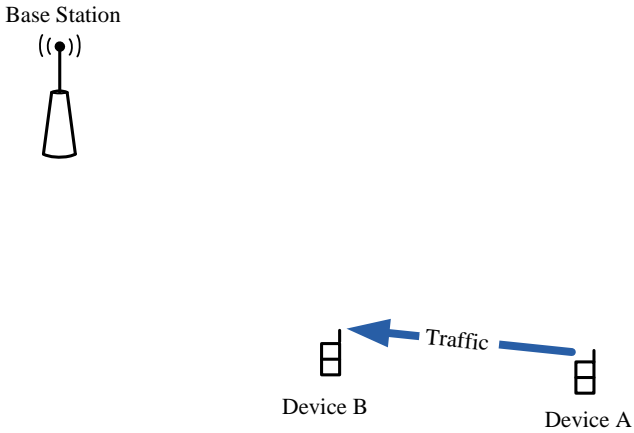

Fig. 1. P2P communication between devices A and B in a cellular system.

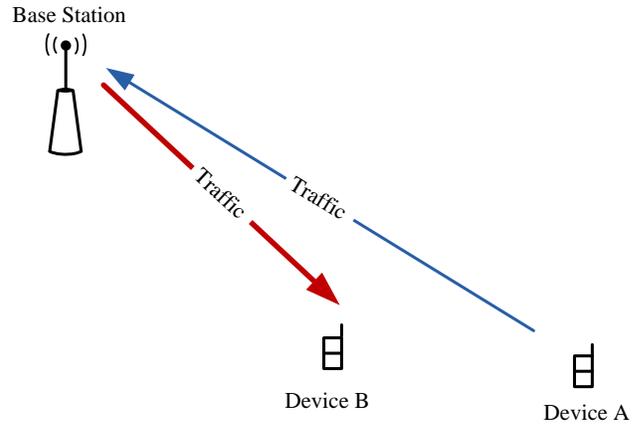

Fig. 2. Conventional WAN communication between devices A and B in a cellular system, where the base station serves as a relay. Red: Downlink; Blue: Uplink.

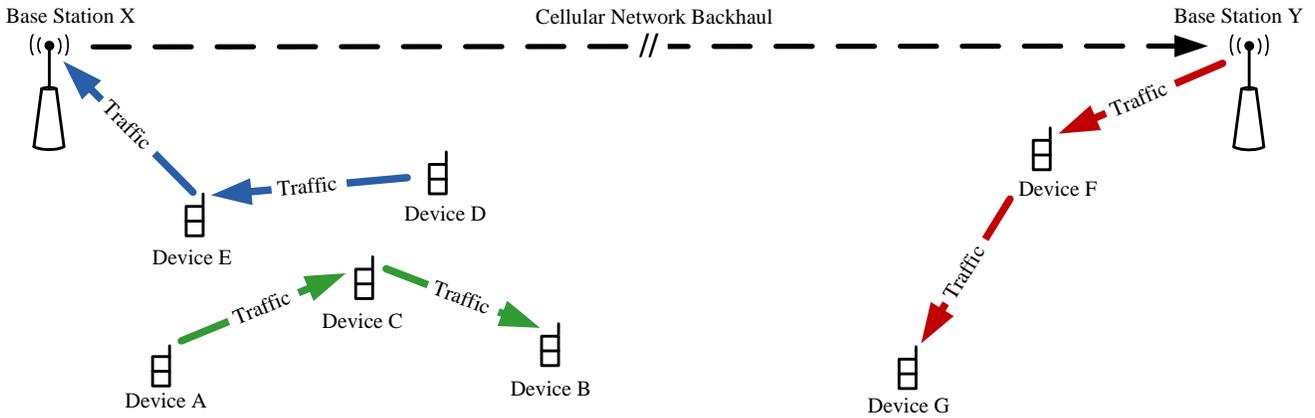

Fig. 3. Illustration of multi-hop P2P communication as an overlay to a cellular network.

centralized location servers locate the geographic locations of all devices. The network-based peer discovery is adequate provided that the geographic location can be determined accurately and the information is up-to-date. Unfortunately, the determination of the location of *mobile* devices is often complex and of low accuracy. Plus, the geographic proximity sometimes may not overlap with the RF proximity, e.g., two devices separated by a thick brick wall are geographically neighbors but may not be neighbors in an RF sense. In RF-based discovery, each device transmits a discovery signal on a designated resource and monitors other devices' discovery signals to determine RF proximity. A device can thus find its neighbors by detecting discovery signals on the designated resource. RF-based discovery is clearly more efficient and therefore the focus of our design.

### A. Peer Discovery Base Model

This section provides a brief review of the framework that constitutes the foundation for neighbor discovery of most existing P2P networks [18][19].

The goal of the neighbor discovery process is to have each mobile device in the network discover all its immediate neighbors (the devices with which it can establish a direct wireless communication link). Assume all devices in the neighborhood follow a common slotted channel structure as illustrated in Fig. 4. One of every $T$ slots is designated for neighbor discovery. Typically $T$ is large to ensure low duty cycle. Devices who are only interested in discovery can go to sleep between discovery periods for power saving. In a typical neighbor discovery scheme, a device transmits its discovery signal in the discovery period with probability $p$ and listens for transmissions from other devices with probability $1-p$. Collisions occur if a device simultaneously receives transmissions from two or more of its neighbors. Under the assumption that transmissions of discovery signals among devices are independent, the probability that device $i$ discovers its neighbor device $j$ out of $G$ total neighboring devices within $t$ discovery periods is given by

$$P_{ij} = 1 - \left(1 - p(1-p)^{G-1}\right)^t, \qquad \forall i, j \tag{1}$$

The upper bound of the above probability

$$\max_p P_{ij} = 1 - \left(1 - G^{-1}\left(1 - G^{-1}\right)^{G-1}\right)^t \tag{2}$$

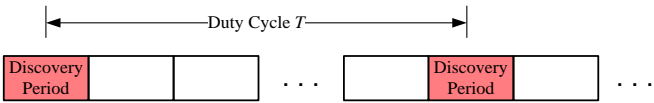

Fig. 4. Illustration of the discovery period.

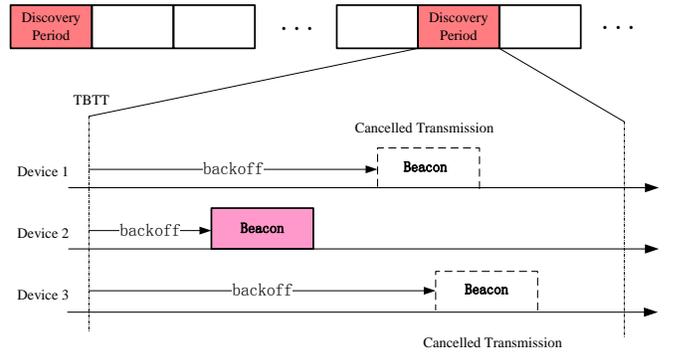

Fig. 5. Beacon transmission using random backoff in IEEE 802.11.

becomes small for large $G$. Hence, the discovery time of a device can be very long. Large $G$ also causes more collisions, and, therefore, collisions result in many wasteful transmissions of discovery signals.

This baseline approach has two drawbacks for use in cellular networks: Firstly it requires a reserved resource (i.e., the discovery period) *exclusively* for the use of discovery signal transmission to avoid collisions with other signals (e.g., data traffic). Secondly, even with the designated resource, collision may still occur among discovery signals from devices, causing delay and waste of resource and energy.

### B. IEEE 802.11/WiFi Beacon Signaling

In IEEE 802.11, a random backoff scheme is used to reduce the probability of beacon signal collision [20]. As shown in Fig. 5, the discovery period starts at the target beacon transmission time (TBTT). Each device, or IEEE 802.11 station, generates a random delay time uniformly distributed in the range that is twice the size of the minimum contention window (CW) and starts to sense the channel for potential beacon signals from other devices. If the channel remains idle before the timer expires, the device transmits its own beacon. However, if a beacon frame is received before the timer expires; device cancels the transmission of its own beacon and waits for the next discovery opportunity/period which is typically 100 ms apart.

Note that the larger the CW is, the less likely two or more beacons collide with each other, though more resources are required.

### C. Direct-Sequence Spread Signaling

Another potential signaling technique for discovery is direct-sequence spread signaling. The direct-sequence spread signal and its variants are commonly used in uplink random access channels (RACHs) either in a CDMA system (e.g., WCDMA [21]) or an orthogonal frequency division multiple access (OFDMA) system (e.g., LTE [22], WiMAX[23]). This random access signal (i.e., the RACH signal) is used by users to request initial network access from its serving base station. In both WCDMA and LTE systems, a contiguous chunk of uplink time-frequency resources are allocated for RACH signals. The RACH signal consists of a complex sequence that is time-and-frequency spread over all the RACH resource (e.g., 1.08 MHz by 1, 2 or 3 ms in LTE) shared by all users. Pseudo-Noise (PN) based sequences are used in WCDMA. In LTE, prime-length Zadoff-Chu (ZC) sequences have been chosen for improved orthogonality between RACH signals[24]. The orthogonality allows simultaneous transmissions of multiple RACH signals without interfering each other. The random backoff scheme used in IEEE 802.11 beacon signaling can thus be avoided. Unfortunately, the orthogonality is lost in the presence of frequency offset due to the accumulated frequency uncertainties at both user transmitter and base station receiver as well as the Doppler shift/spread, resulting in inter-user signal interference and the well known near-far effect [25]-[27].

To minimize the near-far effect, the transmit power must be carefully managed. The user first transmits the signal on the RACH resource with the minimum transmit power and gradually increases its transmit power at each failed attempt (fails to be detected by the receiver, i.e., the base station) in order to find the minimum power to compensate for the path-loss to avoid blanking out other access signals from different users within the same cell. The RACH signal and the trial-and-error power ramping scheme is clearly not applicable to the discovery signaling application for the following reasons: First, the discovery signal is a broadcast signal. The user has no way of knowing if all the intended (neighboring) devices have received the signal. Plus, even the minimum transmit power for the farthest intended device will be overwhelming for the nearest devices, causing them to be unable to receive the discovery signals from other devices. Second, since the recipients of the discovery signal are the neighboring devices, high power is typically required to transmit the discovery signal to ensure sufficient coverage which inevitably blocks its nearest devices from hearing the discovery signals from devices farther away. Third, since the target receivers of the discovery signal are multiple neighboring devices that may not be perfectly synchronized with each other due to different relative Doppler speeds and different errors of synchronization to the base station, the frequency offset between the devices would thus be greater than in the original RACH application, thereby resulting in severe impairment to the orthogonality of the ZC sequences and consequently a more pronounced near-far effect.

### D. Discovery Signal Design for P2P in a Cellular Network

In the proposed design, a temporary discovery ID (TDID) is first created for identifying the devices in a cell and is used for all P2P communications among devices. In the proposed scheme, each discovery period is divided into $D$ parallel "channels", each of which uniquely represents a TDID. A device broadcasts its presence and its associated TDID by energizing one of the TDID channels.

To create the $D$ parallel channels for discovery signaling, the discovery period is first divided into multiple OFDM symbols. The period corresponds to one subframe per frame ($T = 10$ subframes) in the LTE framework [36]. For ease of explanation, we adopt the LTE framework for illustration purpose in the following discussion. All energy in an OFDM symbol is transmitted on a *single* OFDM subcarrier. No energy is transmitted on any other subcarriers of the current OFDM symbol. The energized subcarrier is referred to as the discovery signal *tone*. No information is modulated onto the tone (i.e. neither amplitude nor phase is modulated), which negates the need for channel estimation at the receiver and lowers the dynamic range and phase noise requirements for the PA (power amplifier) and RF components at the transmitter. It is the location (subcarrier index) of the tone that contains information. That is, the index of the energized subcarrier of the OFDM symbol depends on the content of the message. In the current application, the message contains TDID and is denoted as $m$, which is further represented by $K$ information symbols, $\mathbf{u} = (u_1, u_2, ..., u_K)$ for $1 \le k \le K$, or more precisely,

$$m(D) = u_K D^{K-1} + u_{K-1} D^{K-2} + ... + u_2 D + u_1 \qquad (3)$$

where the base $D$ is the total number of subcarriers used for transmitting one information symbol $u_k$ ($0 \le u_k \le D-1$). For a system with $S$ available OFDM subcarriers for data transmission, we assume $D \le S$. We thus need $K$ OFDM symbols to transmit the message $m$, as shown in Fig. 6 where the subcarriers used for discovery signal transmission are numbered from 0 to $D-1$.

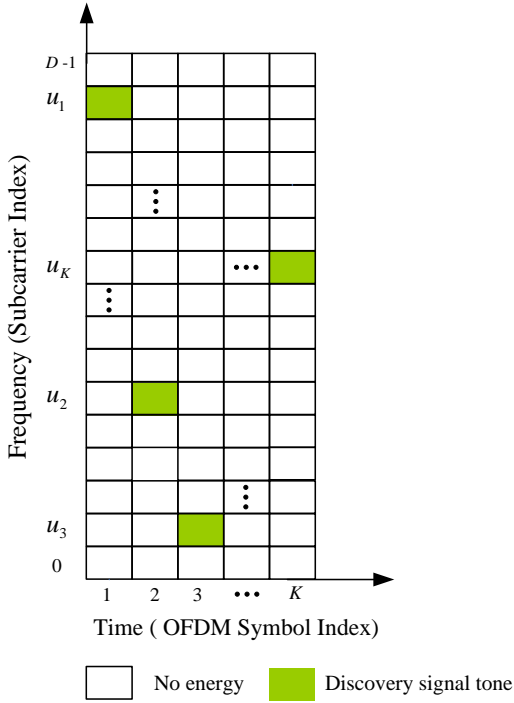

Fig. 6. Illustration of a discovery signal where only the subcarriers used for discovery signal transmission are drawn and are numbered from 0 to $D-1$. The OFDM subcarrier index of each discovery signal tone equals the value of $u_1$ ($1 \le k \le K$). Note that the tone itself is not modulated. Also note that only one single subcarrier is energized per OFDM symbol of a discovery signal.

Note that this single-tone signal is not to be confused with a frequency hopped signal. In frequency hopping, the tone positions are predetermined by a sequence known to both the transmitter and the receiver. The tone position itself therefore does not contain any information and the information is modulated onto the amplitude and/or phase of the tone via QPSK or QAM (pilot/channel estimation is thus needed). For example, in the FlashLinQ social network system [28], the discovery information is first convolutionally coded (binary coding with code rate 1/2) and QPSK-modulated onto the phase of the tone. The discovery signal is then transmitted on a single subcarrier and stays on the same subcarrier for the whole discovery signal duration. Although the subcarrier for the transmission of a discovery signal may change from one discovery period to another discovery period, the hopping pattern that is irrelevant to the content of the discovery signal. This type of hopping is thus solely for the purpose of robustness against frequency selective fading and potential collision deadlock. In this sense, FlashLinQ discovery signals are no different than the regular OFDM signals except that a single tone, instead of multiple tones, is used for the benefit of power efficiency and hence the extension of transmit range. While for the single-tone scheme on the other hand, the tone is not modulated with information. Instead the information is embedded in the positions/subcarrier indices of the tones (no need for channel estimation).

The choice of this type of signal has the following advantage: Unlike the commonly used CDMA signals for multiple access [29]-[37], the single-tone signal does not suffer from the near-far effect among transmissions from different devices. This is because: 1) If some of the tones from different devices happen to transmit on the same subcarrier (e.g., $u_2 = v_2$ in Fig. 7), they simply add together just like multi-path (since the tones are not modulated, they share the same waveform) and are absorbed by the cyclic prefix [38]; 2) If the single-tone signals from different devices are transmitted on different subcarriers, they don't present interference to each other since all the OFDM subcarriers are orthogonal when the frequency between transceivers are perfectly synchronized. If in practice the frequency offset is present between the transceivers, the OFDM subcarriers are no longer orthogonal. However, for the proposed single-tone signals, the tones are not signal-distinctive. The tone energy leakage from the neighbor subcarriers of different signals from different devices to the current subcarrier as a result of non-orthogonality adds together with the tone of the current subcarrier just like multi-path similar to the case in 1). This is not the case though for *regular* OFDM signals where the information modulated on the subcarriers from weak devices can be destroyed by the leakage from neighboring subcarriers from the strong devices causing a near-far effect when the frequency offset is non-zero. That is, regular OFDM signals still suffer from the near-far effect [39].

The significance of this unique property of freedom from near-far effect is that the discovery signal tones from different devices do not interfere with each other. Strong discovery signals will thus not block weak discovery signals. This also means that a device can *listen for other discovery signals while transmitting its own*. This full-duplexing operation is a feature that most existing discovery/beacon signals [12]-[15] lack.

However, although the indistinctiveness nature of the signal tones prevents the discovery signal tones from interfering between different discovery signals, it causes ambiguity among different discovery signals at a receiver. As illustrated in Fig. 7, it would be difficult (if not impossible) for the receiver to figure out which set of tones comprises the original discovery signal tone sequence sent by a device since the individual tones themselves do not identify the discovery signal (although in Fig. 7 different discovery signals from different devices are identified by color simply for illustration purpose). Using again the example in Fig. 7, the receiver may interpret one of the received discovery signal as $\begin{bmatrix} v_1 & u_2 & v_3 & u_4 & \ldots & u_K \end{bmatrix}^T$ or any such combination, where $\begin{bmatrix} u_1 & u_2 & \ldots & u_K \end{bmatrix}^T$ and $\begin{bmatrix} v_1 & v_2 & \ldots & v_K \end{bmatrix}^T$ are the original messages from device 1 and device 2, respectively. The possible combinations are $G^K$ where $G = |\mathbf{G}|$ is the number of simultaneous transmitting devices ($d = 2$ in Fig. 7) and $\mathbf{G}$ is the set of simultaneous transmitting devices.

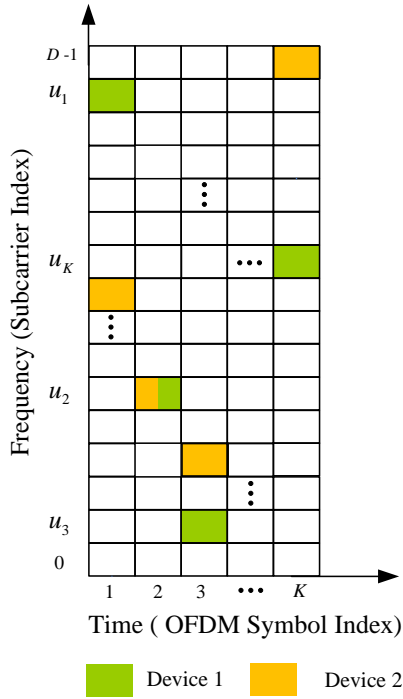

Fig. 7. Illustration of two discovery signals from two devices. The OFDM subcarrier index of each discovery signal tone equals the value of $u_k$ ($1 \leq k \leq K$). The tones themselves are not modulated. They are hence not distinctive between discovery signals sent from different devices although the tones are colored differently in the diagram for illustration purpose. The index for device 2, $v_k$ ($1 \leq k \leq K$), is not shown.

| | Device 1 | | Device 2 |

We therefore devise a special transform in a Galois field, $GF(D)$, such that the transformed STS signal possesses an ambiguity resolution capability and a certain degree of error protection

$$\mathbf{c}^T = \mathbf{Z} \begin{bmatrix} 0 & \mathbf{u}^T & 0 & \cdots & 0 \end{bmatrix}^T \qquad (4)$$

where

$$\mathbf{Z} = \begin{bmatrix} 1 & 1 & \cdots & 1 \\ 1 & \alpha^{\frac{D-1}{N}} & & \alpha^{\frac{D-1}{N}(N-1)} \\ \vdots & & \ddots & \vdots \\ 1 & \alpha^{\frac{D-1}{N}(N-1)} & \cdots & \alpha^{\frac{D-1}{N}(N-1)(N-1)} \end{bmatrix}. \qquad (5)$$

Here $\alpha$ is a primitive number in $GF(D)$, $0 \leq c_n \leq D-1$ for $1 \leq n \leq N$ and $N \geq K$.

One can also think of (4) as a special non-binary transform in $GF(D)$ that that transforms the subcarrier indices of the STS tones, i.e., the original non-binary information symbols $\mathbf{u} = \begin{bmatrix} u_1 & u_2 & \ldots & u_K \end{bmatrix}^T$ into a code word $\mathbf{c} = \begin{bmatrix} c_1 & c_2 & \ldots & c_N \end{bmatrix}^T$ with code rate $(N, K)$. The leading zero element before $\mathbf{u}^T$ in (4) is inserted *on purpose* for frequency error protection as will be explained later. It can be shown that the resulting coded discovery signal $\mathbf{c} = \begin{bmatrix} c_1 & c_2 & \ldots & c_N \end{bmatrix}^T$ achieves the largest possible code minimum distance for any linear code with the same encoder input and output block lengths (or maximum distance separable).

*Proposition 1: The coded discovery signal is maximum distance separable.*

*Proof:* From (4) each code symbol of code word $\mathbf{c}$ can be written as

$$\begin{aligned} c_n &= \sum_{k=1}^{K} u_k \alpha^{\frac{D-1}{N}(n-1)k} \\ &= \alpha^{\frac{D-1}{N}(n-1)} \sum_{k=1}^{K} u_k \left( \alpha^{\frac{D-1}{N}(n-1)} \right)^{k-1}, \quad 1 \leq n \leq N. \end{aligned} \qquad (6)$$

Since a polynomial of order at most $K-1$,

$$\sum_{k=1}^{K} u_k x^{k-1}, \qquad (7)$$

can have at most $K-1$ zeros, the number of non-zero-valued code symbols in $\mathbf{c} = \begin{bmatrix} c_1 & c_2 & \ldots & c_N \end{bmatrix}^T$ is thus at least $N-K+1$. This means that the minimum distance of $\mathbf{c}$ is

$$d_{\min} \geq N-K+1. \qquad (8)$$

However, by the Singleton bound, the minimum distance for any linear code satisfies

$$d_{\min} \leq N-K+1. \qquad (9)$$

It can be readily observed from (8) and (9) that the minimum distance of $\mathbf{c}$ must be

$$d_{\min} = N-K+1. \qquad (10)$$

That is, $\mathbf{c}$ achieves the largest possible code minimum distance for any linear code. We then conclude that the discovery signal is maximum distance separable. ∎

The benefit of the maximum distance separable property of the discovery signal will be seen in the following discussions.

Fig. 8 is an example of the coded discovery signals, where two ($G = 2$) discovery signals of code rate (11, 2) in $GF(32)$,

$$\mathbf{c} = \begin{bmatrix} 18 & 25 & 7 & 2 & 30 & 21 & 27 & 25 & 23 & 5 & 12 \end{bmatrix}^T, \qquad (11)$$

and

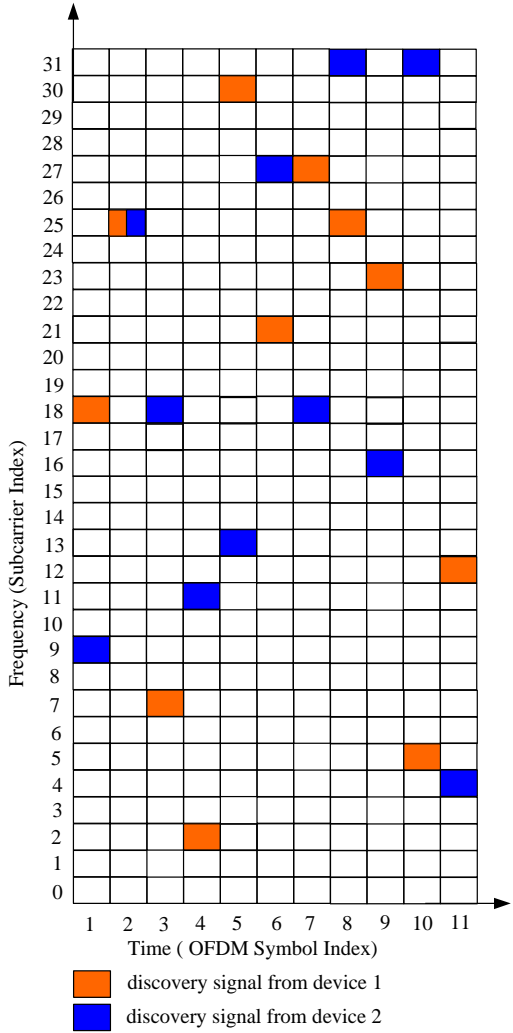

Fig. 8. An example of detecting two coded discovery signals with code rate $(11, 2)$ in $GF(32)$.

$$\mathbf{d} = \begin{bmatrix} 9 & 25 & 18 & 11 & 13 & 27 & 18 & 31 & 16 & 31 & 4 \end{bmatrix}^T, \quad (12)$$

are transmitted, carrying information symbols $\mathbf{u} = \begin{bmatrix} 5 & 1 \end{bmatrix}^T$ and $\mathbf{v} = \begin{bmatrix} 10 & 5 \end{bmatrix}^T$ corresponding to the original 10-bit TDID 100101 and 10101010, from device 1 and device 2, respectively, and received by another device (e.g., device 3). Since the discovery signal tones are indistinctive, a device is not able to tell which tones belong to which discovery signal. For example, the receiver cannot tell if tone 18 belongs to the discovery signal $\mathbf{c}$ in (11) or to $\mathbf{d}$ in (12) and likewise for tone 9 on OFDM symbol 1. Similarly, the receiver cannot tell if tone 21 belongs to $\mathbf{c}$ or to $\mathbf{d}$ (and likewise for tone 27) on OFDM symbol 6. It thus seems that the receiver can arbitrarily interpret the received signals as

$$\mathbf{c}' = \begin{bmatrix} 9 & 25 & 7 & 2 & 30 & 27 & 27 & 25 & 23 & 5 & 12 \end{bmatrix}^T, \quad (13)$$

and

$$\mathbf{d}' = \begin{bmatrix} 18 & 25 & 18 & 11 & 13 & 21 & 18 & 31 & 16 & 31 & 4 \end{bmatrix}^T. \quad (14)$$

We will show in the sequel that the coded discovery signals from different devices can be *disambiguated* at the receiver such that the tone sequences of a coded discovery signal, such as (13) or (14) or any combination of such, can be identified as *invalid* sequences.

Indeed, $G = |\mathbf{G}|$ discovery signals with code rate $(N, K)$ can coexist without causing decoding ambiguity as long as the inequality $K \leq \left\lceil \dfrac{N}{G} \right\rceil$ is satisfied. This important property can be formally stated by the following proposition:

*Proposition 2:* Assume $G$ ( $G \leq D^K$ ) *distinctive discovery signals coded on* $GF(D)$ *with code rate* $(N, K)$ *,* $N \geq K$ *, are simultaneously received on the same time and frequency resource. Under perfect tone detection, all $G$ coded discovery signals can be decoded to the original information symbols without ambiguity, if*

$$K \leq \left\lceil \frac{N}{G} \right\rceil \quad (15)$$

*is satisfied.*

*Proof:* Consider $G$ $\left( G \leq D^K \right)$ distinctive discovery signals coded with rate $(N, K)$ on $GF(D)$ are simultaneously received on $N$ OFDM symbols, free of tone detection errors. Now arbitrarily select $N$ number of the detected tones, each from one of the $N$ different OFDM symbols. We maintain that

1) There are at least $\left\lceil \dfrac{N}{G} \right\rceil$ discovery tones out of the $N$ selected tones coming from the same discovery signal among the total number of $d$ discovery signals. This is the direct outcome of the *pigeonhole* principle [41].

2) For a discovery signal with code rate of $(N, K)$ , a minimum number of $K$ discovery tones are sufficient to distinguish one discovery signal from another. This is sustained by the fact from Proposition 1 that the discovery signals are maximum distance separable. Indeed, since the minimum distance between discovery signals is $d_{\min} = N - K + 1$ (cf., (10)), i.e., the number of different tones between any two discovery signals is at least $d_{\min}$ , or the number of same tones between any two discovery signals is at most $N - d_{\min} = K - 1$ . Therefore a minimum number of $K$ discovery tones are sufficient to determine a discovery signal.

From 1) and 2), it is clear that among the $N$ *discovery* tones, each selected from the $N$ individual received OFDM symbols, there are at least $\left\lceil \dfrac{N}{G} \right\rceil$ tones belonging to one of the $G$ discovery signals. With these $\left\lceil \dfrac{N}{G} \right\rceil \geq K$ tones, we can uniquely determine the corresponding discovery signal. We therefore conclude that if $\left\lceil \dfrac{N}{G} \right\rceil \geq K$ , i.e., (15), is satisfied, the $G$ discovery signals can be uniquely separated from each other without ambiguity. That is, the receiver will not falsely detect any discovery signals other than the $G$ discovery signals. ∎

Regarding the example in Fig. 8, out of the total $2^{11} = 2048$ possible combinations, Proposition 2 guarantees that only two are valid discovery signals. They are **c** in (11) and **d** in (12). All others, such as **c'** and **d'** in (13) and (14), are invalid discovery signals. Therefore, there is no ambiguity in detecting **c** and **d** at the receiver.

It is easy to verify that at least $\left\lceil \dfrac{11}{2} \right\rceil = 6$ out of the 11 tones in any of the 2048 combinations belong to either **c** or **d**. According to the maximum distance separable property of the discovery signal, two tones are sufficient to determine a discovery signal. Since $\left\lceil \dfrac{11}{2} \right\rceil = 6 > 2$, none of the 2048 combinations except the two corresponding to discovery signals **c** and **d** forms a valid discovery signal. The $2048 - 2 = 2046$ combinations are nothing but different combinations of **c** and **d**. They cannot be combinations from any other (valid) discovery signals. We can therefore come to the following remark as a result from proposition 2:

*Remark 1:* Assume $G$ $\left(G \leq D^K\right)$ *distinctive discovery signals coded on GF(D) with code rate* $(N, K)$, $N \geq K$, *are simultaneously received without tone detection error. Among the* $G^N$ *possible discovery tone combination sequences, only* $G$ *are valid discovery signals. The rest* $G^N - G$ *tone sequences are simply different combinations from the* $G$ *valid discovery signals. They can neither form any valid discovery signals nor be combinations from any other (valid) discovery signals.*

However, the case of particular interest is $K = 1$. When $K = 1$, (15) holds for any value of G regardless of the value of $N$. We therefore have the following important remark:

*Remark 2:* Assume $G$ ($G \leq D$) *distinctive discovery signals coded on GF(D) with code rate* $(N, 1)$, $\forall N \geq 1$ *are simultaneously received on the same time and frequency resource. Under perfect tone detection, all* $G$ *discovery signals can be decoded to the original information symbols without ambiguity.*

There are a total of $D$ discovery signals representing $D$ TDIDs. The $D$ discovery signals therefore constitute $D$ discovery channels. Two devices transmitting on different discovery channels (therefore different TDIDs) do not collide with each other. For example, (11, 1) coded discovery signals in theory can support up to $D$ number of simultaneous discovery signal transmissions, or, $D$ discovery channels. That is, up to $D$ simultaneously received discovery signals can be faithfully recovered at the receiver. This is true whether the value of $N$ equals 11 or not[1].

However, as alluded to above, this conclusion is only true under the assumption of ideal tone detection. In practical scenarios when *tone* detection is not error-free due to noise and channel fading, the value of $N$ *does* affect the discovery signal's capability of correcting tone detection errors. The relationship between $N$ and the capability of tone error correction and erasure recovery for code rate $(N, K)$ is given by

$$2\varepsilon + \nu \leq N - K \qquad (16)$$

where $\varepsilon$ is the number of falsely detected tones and $\nu$ is the number of missed detection of tones. This is from the property of a maximum distance separable code [40]. Therefore, as long as the number of falses/misses of discovery tone detection is within the capability of the discovery signals governed by the value of $N$, i.e., $2\varepsilon + \nu \leq N - 1$, ambiguities from multiple discovery signals can be eliminated.

---

[1] Note that when $N=K=1$, the discovery signal reduces to the conventional frequency shift keying signal.

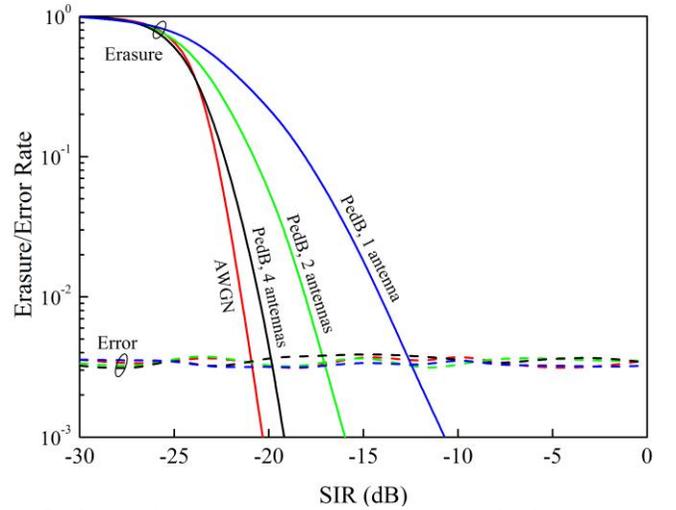

Fig. 9. The decoding erasure and error performance of the discovery signal Total 30 devices; The number of subcarriers in an OFDM symbol = 512, code rate = (11,1), fading speed = 3 km/h at 2 GHz carrier frequency; The number of receive antennas =1, 2 or 4. SIR is defined as OFDM symbol time domain sample SNR. A decoding erasure is defined as the event in which a device fails to decode the discovery signal sent from a device, while a decoding error is an event in which the device decodes the discovery signal to a wrong but valid TDID.

*Propersition 3:* Assume $G$ ($G \leq D$) *distinctive discovery signals coded on GF(D) with code rate* $(N, 1)$, $\forall N \geq 1$ *are simultaneously received on the same time and frequency resource. All* $G$ *discovery signals can be decoded to their original information symbols without ambiguity as long as the number of tone false detections* $\varepsilon$ *and the number of tone miss detections* $\nu$ *satisfy*

$$2\varepsilon + \nu \leq N - 1. \qquad (17)$$

Fig. 9 shows the discovery signal detection baseline performance. The number of simultaneously transmitting devices is 30. A decoding erasure is defined as the event in which a device fails to decode the discovery signal sent from a device, while a decoding error is an event in which the device decodes the discovery signal to a wrong but valid TDID. A decoding erasure prevents the device from recognizing the neighbor whereas a decoding error causes the device to see a false neighboring device. A falsely detected neighbor can be further corrected by the network if the reported neighbor is not an active device or if a device that claims another device to be its neighbor is not on the neighbor list of that device, according to the channel reciprocity principle. In Fig. 9, it is verified that the discovery signal can be detected at very low SNR at a false detection rate below 1%, owing to its unique waveform and coding as well as high degree of frequency diversity.

Another important issue that cannot be overlooked in discovery signal design is sensitivity to time and frequency synchronization errors. Since two transmitting and receiving devices may not be perfectly synchronized, the discovery signal has to be designed with time and frequency offset tolerance. The time offset among devices is easily absorbed by the cyclic prefix of the OFDM symbol as earlier stated and is thereby less of a concern. Frequency offset, due to the accumulated frequency uncertainties at both user transmitter and base station receiver and the Doppler shift/spread, does not cause discovery signal tones to interference with each other as early stated. However, large offset on the other hand may potentially cause the discovery signal tones to shift to the neighboring subcarriers resulting in a potentially different discovery signal. Again, we resort to a special coding technique to provide large frequency offset immunity.

Assume that the transmitted discovery signal is

$$\mathbf{c} = \begin{bmatrix} c_1 & c_2 & \cdots & c_N \end{bmatrix}, \quad (18)$$

and the frequency offset $\Delta f_{ji}$ between the transmit device $i$ and the receiving device $j$ is large enough to cause the received tones to shift to their neighbor subcarriers by $\delta_{ji} \neq 0$ number of subcarriers

$$\mathbf{c}' = \begin{bmatrix} c_1 + \delta_{ji} & c_2 + \delta_{ji} & \cdots & c_N + \delta_{ji} \end{bmatrix}. \quad (19)$$

From (4), the inverse GFT of (18) is given by

$$\mathbf{Z}^{-1}\mathbf{c}^T = \begin{bmatrix} 0 & \mathbf{u}^T & 0 & \cdots & 0 \end{bmatrix}^T \quad (20)$$

with the first element equal to zero. Therefore, the inverse of a *valid* discovery signal always has a zero-valued first element *by design*. Whereas the first element of the inverse transform of (19) produces

$$\frac{1}{N}\sum_{n=1}^{N} c'_n = \frac{1}{N}\sum_{n=1}^{N}\left(c_n + \delta_{ji}\right) = \frac{1}{N}\sum_{n=1}^{N} c_n + \delta_{ji} = \delta_{ji} \neq 0, \quad (21)$$

violating the zero-valued first element property. We therefore have the following propositions:

*Proposition 4: A discovery signal received with a frequency offset does not correspond to a valid (but wrong) discovery signal.*

This property ensures that a receiver with frequency offset will not erroneously map to a discovery signal with a valid TDID.

*Proposition 5: The value of the first element of the inverse transform of a received discovery signal equals the frequency offset between the transmit device and the receive device.*

This property enables the receiver to detect the frequency offset, if any, between the transmitting device and the receiving device. The frequency-offset discovery signal can then be recovered. This property is useful since two communicating devices may not be both perfectly frequency-synchronized to the network.

### E. TDID Acquisition

To transmit its own discovery signal, a device must first acquire a vacant TDID (or a vacant discovery channel) from the pool of $D$ valid TDIDs. There are two ways for TDID allocation: centralized or distributed. In the centralized method TDIDs are jointly managed by a group of cells such that the same TDID is not used more than once in the same RF proximity. The distributed method allows devices to acquire their TDIDs by themselves. In this section, we discuss the distributed method due to its simplicity.

During acquisition, device $i$ scans through the $D$ discovery channels searching for the one with the lowest average energy, i.e., the least congested discovery channel or TDID, over a certain number of discovery periods, i.e.,

$$G^* = \underset{G \in \{1,2,\cdots,D\}}{\arg\min} \sum_j \sum_{n=1}^{N}\left| h_{ji}\left(c_n^G\right) + w \right|^2, \quad (22)$$

where $c_n^G$ is the $G$th discovery signal tone index, $h_{ji}\left(c_n^G\right)$ is the channel gain at subcarrier $c_n^G$ between device $i$ and device $j$, and $w$ is the receive noise power spectral density. To reduce the probability that two or more devices simultaneously acquire the same TDID that has the lowest energy, the device randomly selects a TDID from a set of channels with the lowest energies. Once a TDID is acquired, the device registers the new TDID with the network. The network detects possible TDID collisions and may suggests a different TDID. The device then starts to transmit the corresponding discovery signal.

TDIDs can be spatially reused, i.e., the same TDID can be reused in other cells. However, it is still possible that two or more devices from different cells select the same TDID that can be heard in the current cell. A device therefore occasionally stops transmitting its discovery signal and listens for possible collisions on the selected discovery channel during the discovery period by monitoring the energy level on its discovery channel. The base station also monitors the active discovery channels by using a silent discovery period once in a while for determining if an active TDID has been used by an interfering neighboring cell device. Collision is assumed to be present if significant energy is detected on the active discovery channels during the silent period. This silent period can be predetermined by using a cell-specific "random" pattern (i.e., a pseudo random sequence seeded by the cell ID). The base station can also issue an executive order to force a silent discovery period at any time if necessary. A base station as well as a device makes a decision based on the energy measurement whether a collision has happened and whether the collision energy is significant enough to warrant a re-acquisition of a new TDID.

### F. Spatial Reuse and Resource Management

Since data transmission resources for P2P can be spatially reused, the hidden device problem may occur which is a typical problem in P2P communications [42]. As illustrated in Fig. 10, the data transmissions from device A to device B and from device C to device D may interfere with each other if they are using the same resource since device C cannot hear device A and vice versa but devices B and D can hear both.

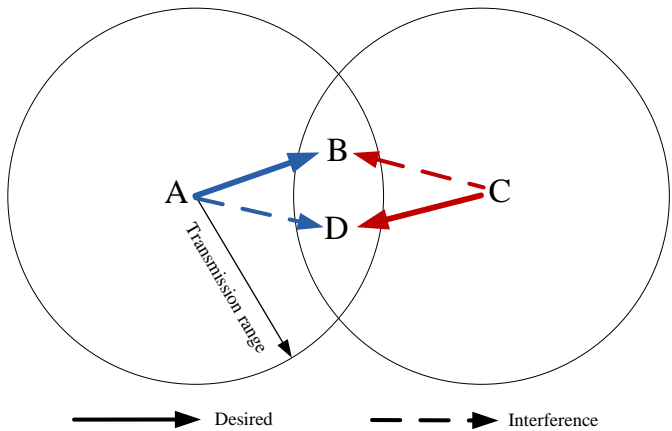

Fig. 10. Illustration of potential hidden device problems between two P2P communication pairs.

Under the proposed architecture, resource allocation for P2P connection is supervised by the network. Using the discovery signal, this problem can then be easily resolved by examining the neighbor list from the receiving devices at the base station. The network allocates orthogonal resources to the two P2P pairs if B and C or A and D are on each other's neighbor list. For the example of Fig. 10, since B and C are on each other's neighbor list and A and D are on each other's neighbor list, indicating B and C, as well as A and D, are in the same RF proximity, orthogonal resources should be allocated to the two P2P communication pairs to avoid interference. It can be seen that spatial reuse to improve spectral efficiency can be realized more effectively and naturally in the proposed cellular structure as a result of peer discovery using discovery signals.

As a special case, if B is D, i.e., a device receives data simultaneously from two devices, the network should allocate orthogonal resources to facilitate the transmission.

Just like a regular cellular connection, the resource scheduling grant for P2P is signaled by the base station to the device pair via downlink control channels.

### G. Multi-hop Relay

Low geometry (e.g., cell edge) devices are typically the bottleneck of overall network performance in a cellular network due to the limited link margin between the device and the base station. However, in the P2P connection mode, the local traffic from a low geometry device can be relayed by its neighbors to the target device. For remote traffic, the data can again be relayed by its neighbors with higher geometries and eventually to the base station. Therefore, the data rate is no longer dependent on the geometry of the device. Higher data rate can thus be sustained for low geometry devices. The network performance is therefore less limited by the low geometry devices.

The hopping path or routing direction is selected by the network. The network uses the neighbor lists as well as the strength of the discovery signal from each device received at the base stations to determine the best hopping path. We will see an example in Section IV.

### H. Discovery Signal Resource Management

A designated resource is typically required for the exclusive transmission of a discovery/beacon signal. This can be costly in a cellular network and may offset the gain from the use of P2P.

For the proposed discovery signal, since the transmit energy is concentrated on one single subcarrier of an OFDM symbol, the discovery signal tone is thus much stronger than a regular data tone, therefore is easily detected even under strong interference. If such a strong received tone is difficult to detect, then the transmitting device is not likely to be in the neighborhood. As a result, discovery signal transmission can overlay other devices' uplink data transmission. On the other hand, the interference of the discovery signal to the uplink traffic is also concentrated on a subcarrier of an OFDM symbol. This isolated interference can effectively be removed by the decoder, which simply erases (punctures) the subcarriers where the corresponding tones from the detected discovery signals are present, thereby causing minimal impact to the other devices' uplink data decoding. In principle, the network can use this resource for either cellular connection (i.e., the base station acts as a relay) or P2P connection (direct communications between two devices). In either case, the device temporarily stops transmitting the discovery signal during this period. However, for the cellular connection, the recipient of the uplink traffic is the base station which has perfect knowledge of the active TDIDs and therefore the locations of the corresponding discovery signal tones. The base station receiver can then more accurately erase the interfered subcarriers. A cellular connection for this resource is thus preferred. The effect of puncturing is the increase of the effective code rate of an uplink traffic. This means that discovery signals can be transmitted *without* reserved uplink resources. Hence the overhead is minimized.

Fig. 11 shows the effect of co-channel transmission of discovery signals on uplink data decoding performance. It is clear that the effect is minimal. However, it is expected that the effect increases as the number of devices increases. Therefore, the network may reuse this resource for the regular cellular uplink traffic and make corresponding adjustments in code rate selection in compensation, if necessary, depending on the number of devices present. Note that since the network has complete knowledge of the active discovery signals, the effective code rate as a result of puncturing can then be pre-determined and taken into consideration in code rate selection of the uplink cellular connection. The base station can choose not to use the discovery slot for traffic if the slot becomes over congested with discovery signals.

### IV. NUMERICAL RESULTS

In this section, we evaluate the performance gain from the use of P2P in a cellular network via the schemes described in the previous sections. The simulation was performed via an LTE network simulator with LTE uplink and downlink fully implemented. There were a total of $S = 512$ OFDM subcarriers per uplink and downlink, and every other of the uplink subcarriers in the discovery period was used for discovery signal transmission (shared with uplink data transmissions), i.e., $D = 256$. Convolutional codes were used for control channels and turbo codes were used for traffic channels. The cellular system consisted of 37 cells with a base station height of 30m and the path loss obeyed the COST231-Hata model. The site-to-site distance was 1000 m. Each base station used a three-sector configuration and each of the sectors was equipped with two antennas. Devices with the default two antennas [37] were randomly and repeated dropped in each cell.

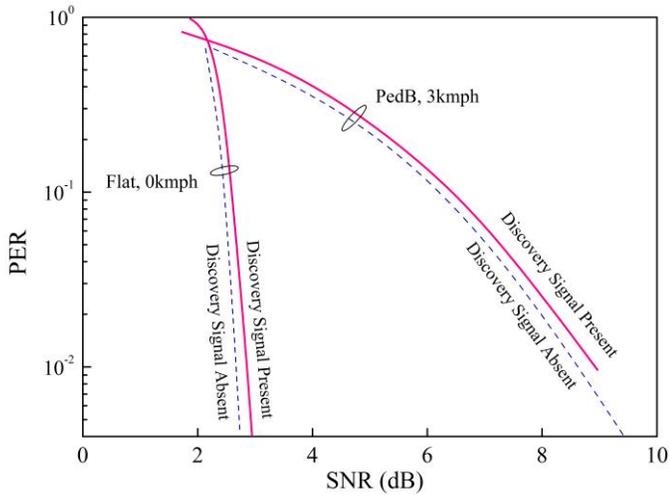

Fig. 11. The effect of co-channel transmission of discovery signals on uplink data decoding performance, where SNR is defined as the receive tone SNR of uplink data per antenna (uplink data modulation=16 QAM, 5 MHz bandwidth, 30 P2P devices randomly dropped per cell).

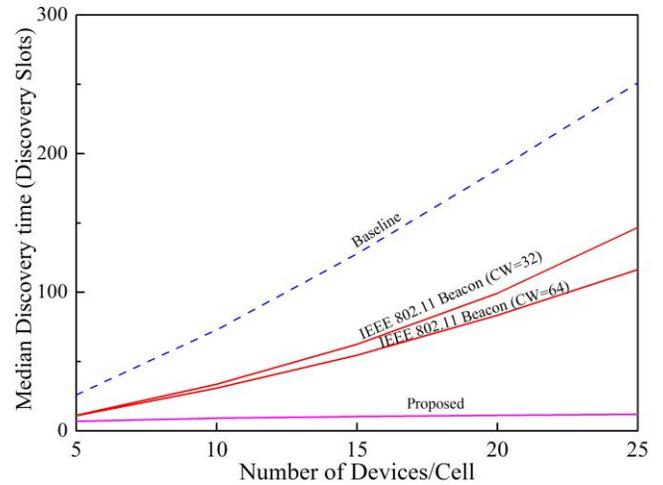

Fig. 12. Device discovery time vs. the density of device drops per cell of radius 250 m. The devices are randomly dropped with various densities. The proposed discovery signal is transmitted on uplink data subcarriers whereas the transmission of the 802.11 beacon signal is on reserved uplink resources.

Discovery signals were coded with code rate $(11, 1)$ using (4). Each of the 11 tones were transmitted on the last 11 OFDM symbols of the $1^{st}$ subframe of a radio frame to avoid the control channels (the first to third OFDM symbols of a subframe were assumed to be used by the base station for transmission of control signals [37]). A device was claimed to be a neighbor if its discovery signal had been consistently detected for four discovery periods ( $4T$ , i.e., four LTE radio frames). The *changes* to the neighbor list, if any, were updated to the network by a device. Note, only changes were reported to save control channel bandwidth. A neighbor list consists of a maximum of 32 neighbors as well as the corresponding link quality indicators (each of which was quantized to three bits). The transmission of the discovery signals as well as the cellular uplink traffic on the discovery period in a cell were temporally suspended in a predetermined cell-specific random pattern for the resolution of conflicting TDIDs amongst neighboring cells. The base station also used this period to detect the strong discovery signals from the neighboring cells in order to remove their tones in later uplink traffic data decoding. Each device maintained a time and frequency estimation loop for synchronization to its serving base station [43].

Throughout the simulation, we assumed that a device had a full buffer of data transmitted either through P2P or relayed by base stations. More simulation parameters can be found in Table I.

Discovery signal tone detection is the first step of discovery signal processing at the receiving device. In the simulations, it was done by simply looking for a subcarrier with significantly higher energy than its neighbors. After the detection of discovery tones, the receiver obtains a set of discovery tones on every OFDM symbol with some falsely detected as well as missed discovery tones. By applying, for example, maximum likelihood decoding using a lookup table, the receiver finally recovers the original information symbols and hence the TDID.

Fig. 12 plots the neighbor device discovery time against the number of devices per cell. The performance of the baseline method (Section II) and the IEEE 802.11 beacon is also shown for reference. As expected, the IEEE 802.11 beacon performance depends on the contention window (CW) size. The larger the CW size is, the better the performance, however, the expense of increased system resource overhead. On the other hand, the performance for the proposed discovery signal remains constant over a wide range of device density. This is expected since the proposed discovery signals do not interfere with each other thereby do not collide with each other or block each other from being detected.

In Fig. 13, one-hop P2P is simulated where devices always had local traffic available for transmission either through direct P2P (one hop P2P connection, c.f., Fig. 1) or being relayed by base stations (cellular connection, c.f., Fig. 2). The choice of P2P connection or cellular connection was simply determined by the relative received strength of the discovery signals between the source and target devices and between the source device and the base station. For comparison, simulations were also performed without P2P, i.e., all the device data were relayed by the base stations. Approximately 25 devices were dropped per cell. They were divided into groups, each of which consists of receiving and transmitting devices. A receiving device is paired with at least one transmitting device in the gtoup (i.e., a device may receive data streams simultaneously from multiple devices). The devices were randomly dropped repeatedly in the simulation.

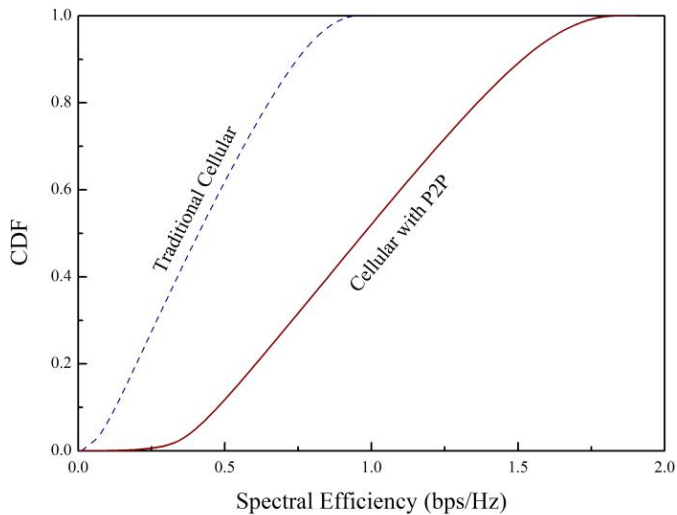

Fig. 13. Spectral efficiency comparison between a traditional cellular network and the cellular network with one-hop P2P.

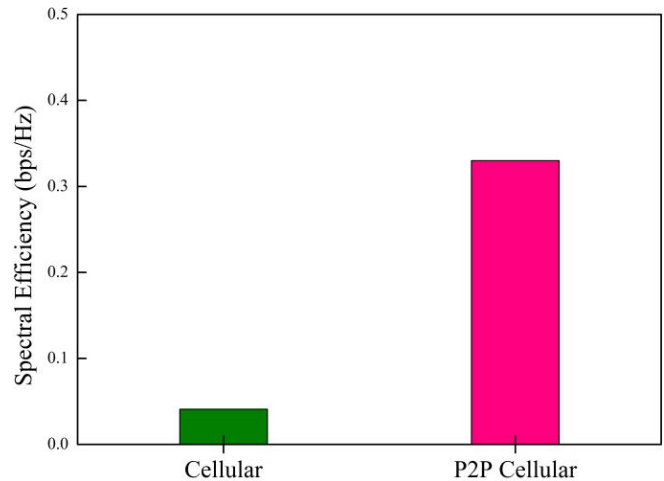

Fig. 14. Effect of two-hop P2P on cell edge device (5% lowest geometry devices) performance.

In Fig. 14, multi-hop P2P is simulated where remote data from disadvantageous devices (low geometries) to the network were relayed by multiple devices (c.f., Fig. 3) to examine the effect of P2P relaying on cell edge performance. For the sake of simplicity, a maximum two-hop relay was considered in the simulation and data were delivered from the device to the base station. The hopping path was determined by the network according to the neighbor lists reported from the devices via cellular uplink control channels as described in the earlier section. For demonstration purpose, a simple "max-min" criterion was used to find the relay device from source device $s$ to a target device $t$

$$r = \arg\max_{i \neq s,t} \max \left\{ 0.5 \min \left\{ \eta_{is}, \eta_{ti} \right\}, \eta_{ts} \right\} \quad (23)$$

where $\eta_{jk}$ is the data rate supported by the link quality from device $k$ to device $j$ measured and reported by device $j$ via the discovery signal transmitted from device $k$. In this simulation scenario, the target device $t$ is always the base station. The cell edge performance improvement is clearly seen as a result of replacing one link-budget constrained link with two stronger P2P links.

## V. CONCLUSION

Peer-to-peer (P2P) communication as an overlay to the existing cellular network has been considered a potential means to allow more devices to communicate with higher data rates through cellular networks in a given cellular bandwidth. This is achieved by establishing collaboration among devices for traffic optimization, better spatial reuse of resource, better coverage and better energy saving. P2P and cellular networks are two fundamentally different architectures. The P2P communication is characterized by high degrees of flexibility, scalability and collaboration among devices, however, also poor resource efficiency and loose QoS constraints. Cellular communications, on the other hand, focus on high resource efficiency, stringent QoS requirements via centralized cellular infrastructure. Applying the fundamentally different P2P communication concept to the cellular infrastructure and benefiting from it is technically challenging and needs to be carefully studied. Previous works mainly focused on the high-level theoretical analysis, providing high-level insights on the benefits and challenges of P2P in cellular networks. In this paper, we provide a practical physical-layer level design. The design takes advantage of P2P collaboration among devices while retaining high spectral efficiency and tight resource and interference control provided by the cellular network.

The heart of the design and the main focus of this paper is the use of a special "light-weight" discovery signal crafted specially for cellular applications. Discovery signals are required to have properties of robustness to interference and synchronization errors as well as energy efficiency and high coverage range. This goal typically requires high resource usage resulting in high overhead which becomes a critical issue for use in the cellular network where the licensed spectrum is scarce and extremely costly. We propose a unique design of a discovery signal that meets these two conflicting goals. First of all, we use the un-modulated single tone structure to prevent interference among different discovery signals from different devices; thereby, eliminating the collision and near-far problems. The discovery signals do not collide with each other nor are they blocked by another discovery signal. A device can even receive discovery signals while transmitting its own discovery signal. The resulting ambiguities of discovery signals at the receiver, due to the indistinctiveness of the tones among different discovery signals, are resolved by the use of a special non-binary transform in a Galois field. The un-modulated single tone structure further negates the need for pilot transmission and channel estimation, greatly simplifying operations at both the transmitter and the receiver. As a result of this unique feature along with low-PAPR[30]-[32], energy efficiency is thus improved. The goal of low signaling overhead is achieved, again, by taking advantage of the single tone property of the proposed discovery signal, i.e., *all* the OFDM symbol energy is concentrated on a *single* subcarrier. The highly concentrated energy not only makes the discovery signal tones easily detectable even under high interference *from* co-channel uplink traffic but also on the other hand increases the isolation of their interference *to* co-channel uplink data traffic making them easily removed at the base station by puncturing the OFDM subcarriers containing the discovery signal tones. Since the base station knows the location of these tones, the discovery signal tones can be cleanly punctured without causing interference to uplink data decoding. The net effect of puncturing on the uplink traffic is the increase of the effective code rate and can be compensated by appropriate code rate adjustment in scheduling, if necessary. As a result, the interference to the uplink data traffic can be completely controlled by the network. This property allows the discovery signal to share resources with uplink data traffic without the need for designated resources, which is difficult to achieve for the conventional signaling techniques as the signal energy is spread over the entire transmission band.

Table I. Simulation Parameters

| | |
|---|---|
| System bandwidth | 5MHz |
| Base station / Device transmit power | 43 / 23dBm |
| Site to site distance | 1000m |
| Carrier frequency | 2GHz |
| Number of subcarriers / Interval | 512 / 15kHz |
| Cyclic prefix duration | 4μs |
| Transmission frame length | 10ms |
| OFDM symbols per frame | 14 |
| Tx / Rx antennas | 2 / 2 |
| Path loss model | COST 231-Hata |
| Channel model | Space Channel Model[44] |
| Thermal noise density | -174dBm/Hz |
| Receiver noise figure (Base station / Device) | 6 / 10dB |
| Number of devices per sector | 25 |

Throughout the design, we leverage the use of the special discovery signal to minimize the P2P operation overhead. Based on the information provided by discovery signals, the network is able to build a blueprint of the RF relationship among devices such that resources, interference, P2P/cellular connection selection, and multi-hop routing can be efficiently managed to ensure QoS under a unified cellular network centric architecture.